\documentclass[reprint, amsmath,amssymb, aip, apl]{revtex4-2}
\usepackage{graphicx}
\usepackage{amsmath}
\usepackage{amssymb}
\usepackage{bm}
\usepackage{physics}
\usepackage{comment}
\usepackage{textcomp}
\usepackage{times}
\usepackage{color}
\usepackage{soul}

\begin{document}

\preprint{AIP/123-QED}

\title{Magnetic field imaging by hBN quantum sensor nanoarray}

\author{Kento Sasaki}
\email{kento.sasaki@phys.s.u-tokyo.ac.jp}
\affiliation{Department of Physics, The University of Tokyo, 7-3-1 Hongo, Bunkyo-ku, Tokyo 113-0033, Japan}
\author{Yuki Nakamura}
\affiliation{Department of Physics, The University of Tokyo, 7-3-1 Hongo, Bunkyo-ku, Tokyo 113-0033, Japan}
\author{Hao Gu}
\affiliation{Department of Physics, The University of Tokyo, 7-3-1 Hongo, Bunkyo-ku, Tokyo 113-0033, Japan}
\author{Moeta Tsukamoto}
\affiliation{Department of Physics, The University of Tokyo, 7-3-1 Hongo, Bunkyo-ku, Tokyo 113-0033, Japan}
\author{Shu Nakaharai}
\affiliation{International Center for Materials Nanoarchitectonics, National Institute for Materials Science, 1-1 Namiki, Tsukuba, Ibaraki 305-0044, Japan}
\affiliation{Department of Electrical and Electronic Engineering, Tokyo University of Technology, 1404-1 Katakuramachi, Hachioji, Tokyo 192-0982, Japan}
\author{Takuya Iwasaki}
\affiliation{International Center for Materials Nanoarchitectonics, National Institute for Materials Science, 1-1 Namiki, Tsukuba, Ibaraki 305-0044, Japan}
\author{Kenji Watanabe}
\affiliation{Research Center for Functional Materials, National Institute for Materials Science, 1-1 Namiki, Tsukuba, Ibaraki 305-0044, Japan}
\author{Takashi Taniguchi}
\affiliation{International Center for Materials Nanoarchitectonics, National Institute for Materials Science, 1-1 Namiki, Tsukuba, Ibaraki 305-0044, Japan}
\author{Shinichi Ogawa}
\affiliation{National Institute of Advanced Industrial Science and Technology, 1-1-1 Umezono, Tsukuba, Ibaraki 305-8568, Japan}
\author{Yukinori Morita}
\affiliation{National Institute of Advanced Industrial Science and Technology, 1-1-1 Umezono, Tsukuba, Ibaraki 305-8568, Japan}
\author{Kensuke Kobayashi}
\email{kensuke@phys.s.u-tokyo.ac.jp}
\affiliation{Department of Physics, The University of Tokyo, 7-3-1 Hongo, Bunkyo-ku, Tokyo 113-0033, Japan}
\affiliation{Institute for Physics of Intelligence, The University of Tokyo, 7-3-1 Hongo, Bunkyo-ku, Tokyo 113-0033, Japan}
\affiliation{Trans-scale Quantum Science Institute, The University of Tokyo, 7-3-1 Hongo, Bunkyo-ku, Tokyo 113-0033, Japan}

\date{\today}

\begin{abstract}
Placing a sensor close to the target at the nano-level is a central challenge in quantum sensing. 
We demonstrate magnetic field imaging with a boron vacancy (V$_\text{B}^-$) defects array in hexagonal boron nitride with a few 10~nm thickness. 
V$_\text{B}^-$ sensor spots with a size of (100~nm)$^2$ are arranged periodically with nanoscale accuracy using a helium ion microscope and attached tightly to a gold wire. 
The sensor array allows us to visualize the magnetic field induced by the current in the straight micro wire with a high spatial resolution. 
Each sensor exhibits a practical sensitivity of $73.6~\mu\text{T/Hz}^{0.5}$, suitable for quantum materials research.
Our technique of arranging V$_\text{B}^-$ quantum sensors periodically and tightly on measurement targets will maximize their potential.
\end{abstract}

\maketitle 

Precise position control of the sensors against targets is key to unlocking their potential.
The nitrogen-vacancy (NV) center in diamond, a representative quantum sensor, can achieve sub-Angstrom resolution by placing it close to the target~\cite{Zopes2018,Sasaki2018}.
To bring the NV center close to diverse targets, researchers have developed techniques to microfabricate a diamond platform~\cite{Rondin2013,Maletinsky2012,Zhou2017,Waxman2014,Schlussel2018,Burek2012,Du2017,Foy2020,Tsukamoto2022} and generate it near the diamond surface~\cite{Fu2010,Ohashi2013,Tetienne2018,Sangtawesin2019,Healey2022a}.
However, in such a three-dimensional platform, developing stable sensors near the surface and accurately determining the actual sensor-target distance are inherently challenging~\cite{Huler2014,Grinolds2014,Pham2016}.
Here, we use a helium ion microscope (HIM) to position quantum sensors in a two-dimensional platform, that are, boron vacancy (V$_\text{B}^-$) defects in hexagonal boron nitride (hBN) precisely in a nanoarray.
Spatially regular patterns allow us to image a magnetic field of a symmetric target with a resolution beyond the diffraction limit.
This arrangement technique can be used to precisely measure symmetrical targets, such as wires and disks, and also point-like targets.

\begin{figure}
\begin{center}
\includegraphics{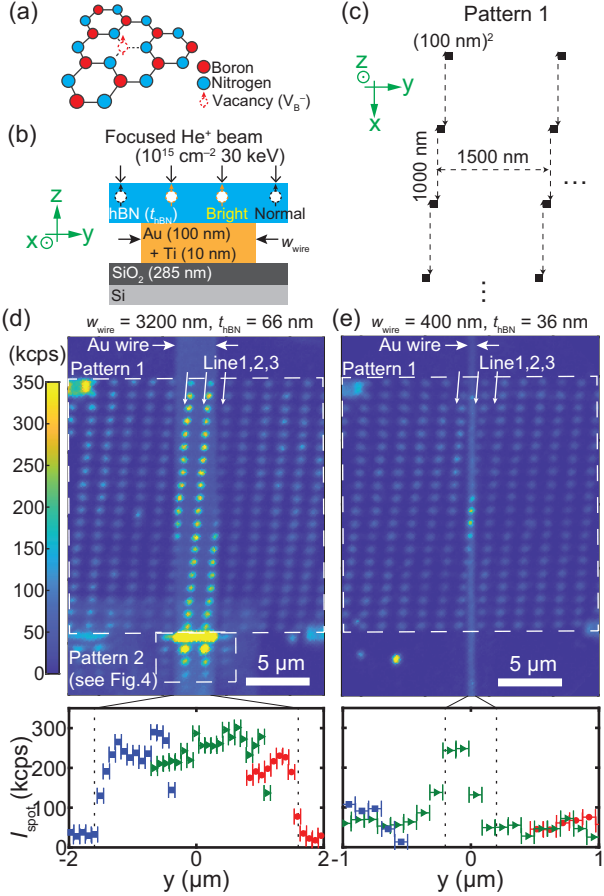}
\caption{
Sensor arrangement using helium ion microscopy.
(a) Structure of a boron vacancy (V$_\text{B}^-$) defect in hBN.
(b) Schematic of our device.
$w_\text{wire}$ is the gold wire width and $t_\text{hBN}$ is the hBN thickness.
(c) Spatially regular irradiation pattern (``Pattern 1'') used for HIM.
(d) Colormap of the PL intensity for the device with $w_\text{wire}=3200$~nm. 
The lower panel shows the PL intensity of the spots on the three lines (Lines 1, 2, and 3 assigned in the upper panel) going across the gold wire.
The squares (blue), triangles (green), and circles (red) represent the result of the spots on Lines 1, 2, and 3, respectively.
The horizontal error bars correspond to the spot size of (100~nm)$^2$.
We explain ``Pattern 2'' in Fig.~\ref{fig4}. 
(e) Counterpart of (d) for the results of the device with $w_\text{wire}=400$~nm.
\label{fig1}
}\end{center}
\end{figure}

V$_\text{B}^-$ in hBN is a defect where a boron atom in hBN is replaced by a vacancy [Fig.~\ref{fig1}(a)].
The electron spin of V$_\text{B}^-$ can be measured by optically detected magnetic resonance (ODMR) even at room temperature, like those in NV centers~\cite{Gottscholl2020}.
They can be used as a sensor for the magnetic field and temperature~\cite{Gottscholl2021,Huang2022,Healey2022,Kumar2022}.
As hBN is a van der Waals (vdW) material, its nanometer-thin flakes can be peeled off with a Scotch tape and tightly attached to the sensing target by a stamping technique. 
The flake thickness can be determined with a stylus profiler, which gives the maximum V$_\text{B}^-$-target distance.
Vacancy-related defects can be generated at targeted locations by focused ion beam (FIB)~\cite{Martin1999,Herzig2021,Klein2019,Mitterreiter2020,Klein2021, Liang2021,Allen2021}, focused electron beam~\cite{McLellan2016,Fournier2021}, laser irradiation~\cite{Wang2021}, and implantation masks~\cite{Toyli2010}. 
Among them, the FIB technique achieves a superior accuracy of about 10~nm~\cite{Yang2014,Mitterreiter2020}. 
Recently, the FIB technique for placing V$_\text{B}^-$ defects using HIM has been reported~\cite{Zeng2022}. 
Unlike NV centers, only He$^+$ irradiation is necessary  for the V$_\text{B}^-$ generation without any heat treatment.
The present work will show that we use HIM to fabricate a V$_\text{B}^-$ quantum sensor nanoarray. 
As proof of principle, we will demonstrate that it can be used for a high resolution imaging of symmetric gold wires and their magnetic fields.

Figure~\ref{fig1}(b) shows the device structure. 
We fabricate a 100~nm thick gold wire on a 10~nm thick titanium on a silicon substrate with a 285~nm thick oxide film. 
We prepare two devices with different wire widths ($w_\text{wire}$) of 3200 and 400~nm. 
hBN flakes peeled off with a Scotch tape are attached to the substrate to cover the wire with the bubble-free transfer method~\cite{Iwasaki2020}. 
Using a stylus profiler (KLA-Tencor D-100), we estimate the thicknesses of hBN ($t_\text{hBN}$) to be 66~nm and 36~nm for the devices with $w_\text{wire}=3200$ and 400~nm, respectively. 
We define the $x$-, $y$-, and $z$- directions as a wire direction, a direction across the wire, and a direction perpendicular to the substrate, respectively, as shown Fig.~\ref{fig1}(b).

We fabricate a 20~$\mu$m~$\times$~20~$\mu$m V$_\text{B}^-$ sensor array on the hBN flake by He$^+$ irradiation using HIM. 
Figure~\ref{fig1}(c) shows the array design, which is a two-dimensional regular spot pattern  (``Pattern~1'').
Each irradiation spot size is designed to be (100~nm)$^2$. 
The spots lie on the slightly tilted lines at 1000 and 100~nm intervals in the $x$- and $y$- directions, respectively. 
The lines lie 1500~nm apart in the $y$-direction. 
Using Orion Plus HIM (Carl Zeiss Microscopy LLC, Peabody, MA, USA), a helium ion beam with a nominal width of 0.3~nm is raster scanned at 7~nm intervals to homogeneously dose in each spot~\cite{Nakaharai2022}. 
We set the acceleration voltage to 30~keV and the average dose for each spot to $10^{15}~\text{cm}^{-2}$. 
The acceleration voltage is the same as in previous studies~\cite{Guo2022,Zeng2022}, and He$^+$ is expected to pass through the hBN flake after producing V$_\text{B}^-$.

We investigate the devices at room temperature using a homemade confocal microscope system~\cite{Misonou2020}. 
The photoluminescence (PL) intensity is measured with a green laser ($\lambda=532$~nm, Typ. 0.7~mW) focused with an objective lens (Magnification = 100, NA = 0.8). 
The diffraction-limited laser spot width is 406~nm ($=0.61\lambda$/NA). 
The long-pass and short-pass filters in this system are set to cutoffs of 750 and 1000~nm, respectively, to reduce PL signals other than that from V$_\text{B}^-$. 
In the ODMR measurement, microwaves (Typ. 1~Watt) are supplied from a $20~\mu$m-diameter copper wire at about $50~\mu$m from the imaging area.

Now, we discuss our experimental results.
Figure~\ref{fig1}(d) shows the PL intensity distribution of the device with $w_\text{wire}=3200$~nm.
The observed PL pattern is consistent with Pattern~1 [Fig.~\ref{fig1}(c)]; the V$_\text{B}^-$ sensors are arranged as designed.
We add that their PL intensities remain sufficiently detectable for at least several months since their creation by HIM.

We focus on the PL intensity from the spots on Lines~1, 2, and 3 crossing the wire, as shown in the upper panel of Fig.~\ref{fig1}(d). 
From now on, we suppose that the PL intensity is independent of the vertical position ($x$) along the wire.
The lower panel of Fig.~\ref{fig1}(d) compiles the result as a function of the transverse location ($y$) in the wire.
The vertical axis $I_\text{spot}$ is the PL intensity after removing the background on the reasonable assumption that the two-dimensional Gaussian distribution in the $xy$-plane can explain the signal from each spot well.
Two vertical dashed lines in the lower panel of Fig.~\ref{fig1}(d) display the edges of the wire obtained by analyzing the PL image. 
The spots on the wire have higher PL intensity than the others; $I_\text{spot}$ is as high as about 200~kcps on the wire and $I_\text{spot}\lesssim 50$~kcps otherwise.
The observed intensity enhancement on gold is consistent with previous studies~\cite{Gao2021,Zeng2022}.
These studies discussed that the cause of this may be plasmonic enhancement around the metal surface~\cite{Fort2007}.
Interestingly, as shown in the panel, the intensity rapidly changes within 200--300 nm (two or three spots) across the wire edges, presumably reflecting the sensitive dependence of the present effect on the amount of gold present in the surroundings.
At the same time, this observation indicates that our local V$_\text{B}^-$ spots have high spatial resolution, i.e., spatially abrupt changes can be detected.

We examine the even narrower device.
Figure~\ref{fig1}(e) shows the PL intensity distribution of the device with $w_\text{wire}=400$~nm.
Again, a PL pattern is consistent with Pattern~1 and the spots on the gold wire are brighter than the others.
As quantitatively shown in the lower panel of Fig.~\ref{fig1}(e), $I_\text{spot}$ reaches its maximum within two spots on the wire and abruptly drops outside.
This is consistent with the position-dependence of the present enhancement effect mentioned above.
The lower panel also suggests that the left side of the wire is brighter than the right side, which might be due to the inhomogeneous adhesion of the hBN flake to the narrow wire.
Curiously, while the hBN thickness of this device ($t_\text{hBN}=36$~nm) is only about half that of the device with $w_\text{wire}=3200$~nm ($t_\text{hBN}=66$~nm) [Fig.~\ref{fig1}(d)], the maximum PL intensity of the spots on the wire is comparable.
The reason for this phenomenon remains to be clarified.

So far, we have established that V$_\text{B}^-$ quantum sensors can be patterned into a nanoarray directly on the target with high spatial resolution beyond the diffraction limit. 
Our observations are consistent with the size of each V$_\text{B}^-$ spot (100~nm)$^2$. 
In our method, the spacing between sensors, i.e., sampling resolution, is restricted by the optical spot size.
By utilizing the symmetry, better spatial/sampling resolution can be achieved for a projection direction while satisfying this restriction~\cite{Sengottuvel2022}.
The enhanced PL intensity suggests that they possess high magnetic field sensitivity, as we will prove next.

\begin{figure}
\begin{center}
\includegraphics{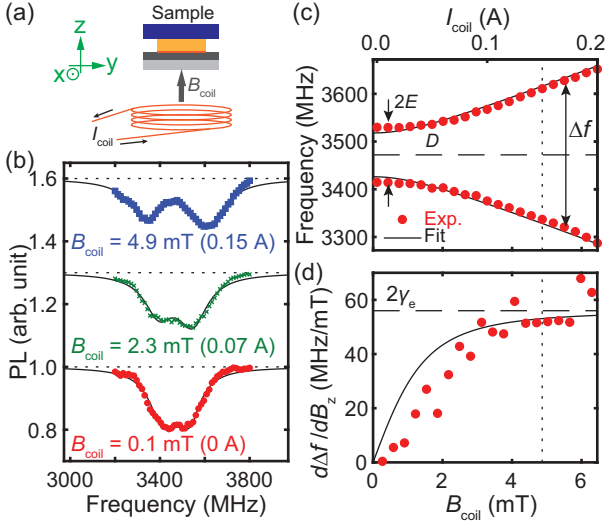}
\caption{
Verification of V$_\text{B}^-$ sensor operation.
(a) Schematic of the coil field application.
(b) Magnetic field dependence of the ODMR spectra. 
For clarity, the spectra are incrementally shifted upward by 0.3.
(c) Field dependence of the resonance frequencies.
(d) Field dependence of the derivative $\frac{d\Delta f}{d B_z}$.
\label{fig2}
}\end{center}
\end{figure}

We perform ODMR measurement of V$_\text{B}^-$ on a gold wire to verify its sensor operation.
V$_\text{B}^-$ is mainly sensitive to the magnetic field parallel to its quantization axis, which is perpendicular to the hBN surface [Fig.~\ref{fig1}(a)].
As shown in Fig.~\ref{fig2}(a), we apply a static magnetic field ($B_\text{coil}$) to the $z$-direction by using a current through a coil placed beneath the device.
We first carefully calibrate the magnetic field.
Using a tesla meter (Lakeshore F71), we obtain the relation between coil current and coil field, $B_\text{coil} = 31.7~(\text{mT/A}) \times I_\text{coil} + 0.1$~mT.

Figure~\ref{fig2}(b) shows the magnetic field dependence of the ODMR spectrum.
The vertical axis is the PL intensity with microwave irradiation, normalized by the PL intensity without microwave irradiation.
The spectrum obtained without coil current ($B_\text{coil}=0.1$~mT) has a dip around the frequency of 3470~MHz.
This dip is a resonance signal of V$_\text{B}^-$'s ground spin triplet state~\cite{Gottscholl2020}.
The ODMR contrast, dip depth, is about 20~\%.
This contrast is similar to Refs.~\onlinecite{Gao2021,Zeng2022} (up to 46~\%) and larger than most previous studies~\cite{Gottscholl2021,Liu2021} (a few percent or less).
We assume that the microwave applied to the gold induces carrier oscillations and generates strong fields nearby~\cite{Mariani2020}, which excites a broad linewidth of V$_\text{B}^-$.
ODMR spectrum can also be obtained at spots outside the gold wire, although the contrast is lower (data not shown).

The cross (green) and square (blue) curves in Fig.~\ref{fig2}(b) are the ODMR spectra obtained at 2.3 and 4.9~mT, respectively.
As the field increases, the spectrum splits into two dips by the Zeeman effect.
The solid lines result from the conventional double Lorentzian fit, which agrees with the experimental results. 
Figure~\ref{fig2}(c) shows the field dependence of the resonance frequencies obtained from the fitting.
We define the higher and lower resonance frequencies as $f_+$ and $f_-$, respectively. 
They are known to follow~\cite{Gottscholl2020}, 
\begin{equation}
f_{\pm} = D \pm \sqrt{ (\gamma_\text{e}B_z)^2 + E^2 },
\label{eq1}
\end{equation}
where $\gamma_\text{e}=28$~MHz/mT is the gyromagnetic ratio, $B_z$ is the magnetic field strength in the $z$-direction, and $E$ is the crystal strain parameter.
The solid lines in Fig.~\ref{fig2}(c) are the fitted results using Eq.~(\ref{eq1}) with $B_z=B_\text{coil}$, nicely reproducing the experimental results.
We obtain $D = 3472$ and $E = 46$~MHz, comparable to the previously reported values~\cite{Guo2022}.

In quantum sensing, the magnetic field is obtained from the change in the resonance frequency difference $\Delta f \equiv f_+ - f_-$.
The derivative $\frac{d\Delta f}{d B_z}$ is an essential value to determine the magnetic field sensitivity.
Figure~\ref{fig2}(d) shows the differential coefficient calculated from Fig.~\ref{fig2}(c).
This derivative is almost zero near zero fields and increases to 56~MHz/mT ($=2\gamma_\text{e}$) by increasing the field.
A transition in the dominant $\Delta f$ component from crystal strain to the Zeeman effect causes this behavior, as expected from Eq.~(\ref{eq1}).
In the 4--6~mT region, a nearly optimal derivative is obtained experimentally and theoretically.
We, thus, fix $B_\text{coil}$ at this bias condition [dotted line in Fig.~\ref{fig2}(d)] for the following demonstration.

\begin{figure}
\begin{center}
\includegraphics{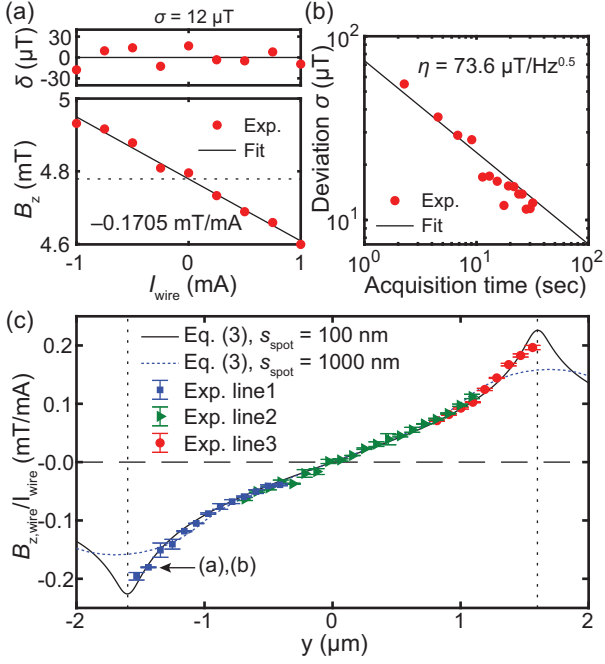}
\caption{
Demonstration of sensing and imaging with quantum sensor nanoarray.
(a) Lower panel shows the $I_\text{wire}$-dependent magnetic field $B_z$ at a spot near the wire edge.
The dotted line corresponds to the coil field (fixed).
The upper panel shows $\delta$, the deviation between the fitted and experimental values.
(b) Acquisition time dependence of the standard deviation $\sigma$.
(c) Oersted field per $I_\text{wire}$ visualized by the nanoarray on the gold wire [Lines 1, 2, and 3 in Fig.~\ref{fig1}(d)].
The left-bottom arrow indicates the spot from which the data in (a) and (b) are obtained.
Error bars on the vertical axis indicate 68~\% confidence intervals of the linear fitting [see (a)].
\label{fig3}
}\end{center}
\end{figure}

We investigate the current-induced Oersted field and evaluate the sensitivity of V$_\text{B}^-$ on the wire.
We apply a current ($I_\text{wire}$) to the gold wire in the $+x$-direction and generate a magnetic field $B_\text{wire}$ that follows Amp\`{e}re's law.
The sum of the $z$ component of $B_\text{wire}$, $B_{z,\text{wire}}$, and the coil field $B_\text{coil}$ results in $B_z=B_{z,\text{wire}}+B_\text{coil}$.
The lower panel of Fig.~\ref{fig3}(a) shows the $I_\text{wire}$ dependence of $B_z$ at the spot near the wire edge.
The $B_z$ varies linearly in the range of $\pm$1~mA.
It means that $B_{z,\text{wire}}$ varies linearly according to Amp\`{e}re's law.
We determine $B_{z,\text{wire}}/I_\text{wire}=-0.1705$~mT/mA and $B_\text{coil}=4.8$~mT from a linear fit (solid line).
The upper panel of Fig.~\ref{fig3}(a) shows the deviation $\delta$ between the experimental results and the linear fit.
$\delta$ seems completely random, and its standard deviation $\sigma$ is as tiny as 12~$\mu$T.
The field changes in tens of $\mu$T produced by sub-mA current changes are precisely detected, proving the high performance of V$_\text{B}^-$ sensor.

Figure~\ref{fig3}(b) shows the acquisition time dependence of the standard deviation $\sigma$.
The acquisition time $t$ is the total photon counting duration to obtain the ODMR spectrum [Fig.~\ref{fig2}(b)].
$\sigma$ decreases as $t$ increases.
Assuming shot noise, this scaling is fitted by,
\begin{equation}
\sigma = \eta t^{-0.5},
\label{eq3}
\end{equation}
where $\eta$ is the minimum detectable field, so-called sensitivity.
The fitting follows the experimental result, and we obtain  $\eta=73.6~\mu$T/Hz$^{0.5}$.
Note that this sensitivity is a practical value, measured including the ODMR spectrum fitting precision, unlike the theoretical estimation presented in previous studies~\cite{Gottscholl2021,Gao2021}.
Our value means that a weak field of 10~$\mu$T can be detected within one minute and 1~$\mu$T within one hour.
Specifically, this performance is sufficient for addressing quantum materials, such as graphene~\cite{Ku2020,Jenkins2022,Healey2022,Palm2022} and vdW magnets~\cite{Healey2022,Huang2022,Kumar2022}.

We then demonstrate high-resolution magnetic field imaging.
We repeat similar current sweep measurements as Fig.~\ref{fig3}(a) at different spots on the wire.
Relying on the translational symmetry along the $x$-direction, we obtain the magnetic field distribution along the $y$-direction at high spatial/sampling resolution~\cite{Sengottuvel2022}.
Figure~\ref{fig3}(c) shows the distribution of $B_{z,\text{wire}}/I_\text{wire}$ with $w_\text{wire}=3200$~nm [Fig.~\ref{fig1}(d)] as a function of $y$. 
We confirm that $B_z$ is zero at the wire center ($y = 0$) and positive (negative) at the right (left) side of the wire.
The black line in Fig.~\ref{fig3}(c) shows the simulation result when the bare Oersted magnetic field $B_{z,\text{Ampere}}$ is averaged over the volume of the sensor.
Specifically, it is given by
\begin{equation}
B_{z,\text{wire}}(y) = \frac{1}{s_\text{spot}t_\text{hBN}}\int_{-\frac{s_\text{spot}}{2}}^{\frac{s_\text{spot}}{2}} \int_0^{t_\text{hBN}} B_{z,\text{Ampere}}(y+s,z') dz' ds.
\label{eq4}
\end{equation}
Here, $s_\text{spot}=100~\text{nm}$ is irradiation spot size.
It usually depends on the optical spot spread when using homogeneously irradiated hBN films and can degrade spatial resolution of the magnetic field (for example, see blue dotted line).
We assume the top of the gold wire as $z=0$ and assume that there is an hBN film in contact with it, and V$_\text{B}^-$ is uniform to hBN thickness.
The simulation is in good agreement with the experimental result.
Precise visualization of magnetic fields is achieved by devising quantum sensor nanoarray patterns.

\begin{figure}
\begin{center}
\includegraphics{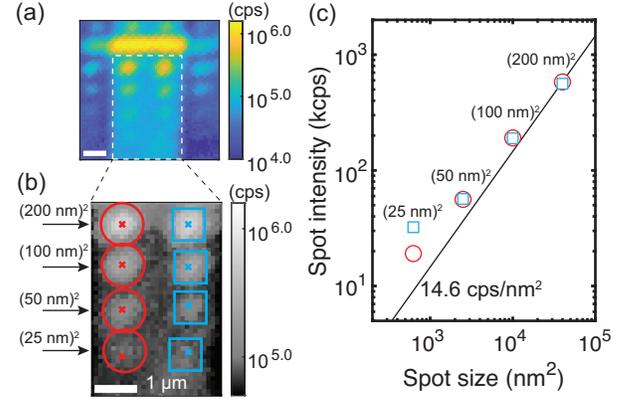}
\caption{
PL intensities of the spots with different sizes.
(a) Enlarged view of Pattern 2 in Fig.~\ref{fig1}(d).
The solid white line is a scale bar indicating 1~$\mu$m.
(b) Further enlarged view of the dashed rectangle in (a).
A logarithmic color scale is used for visibility.
The vertical and horizontal spacings between spots are set at 1575 and 1000~nm, respectively.
The crosses indicate the expected irradiated locations.
(c) Spot size dependence of the PL intensity.
Each marker corresponds to the marker in (b).
\label{fig4}
}\end{center}
\end{figure}

Finally, we examine the spot size dependence of the PL intensity.
Figures~\ref{fig4}(a) and (b) show the intensity at Pattern 2 as shown in Fig.~\ref{fig1}(d).
Pattern 2 consists of two vertical rows whose spacing is 1000~nm.
Four spots whose sizes are (25~nm)$^2$, (50~nm)$^2$, (100~nm)$^2$, and (200~nm)$^2$ are vertically placed 1575~nm apart in each row.
Figure \ref{fig4}(b) shows the four pairs of the two spots of the same size side by side by the circle (red) and square (blue) markers.
The spot positions are aligned as designed in Pattern 2 (marked with crosses).
Figure~\ref{fig4}(c) shows the dependence of $I_\text{spot}$ on the spot size.
As expected, $I_\text{spot}$ increases linearly with the spot area.
The scaling obtained from linear fitting yields 14.6~cps/nm$^2$.
This scaling could be increased to an even higher order of magnitude with increasing the irradiation dose from the present value of $10^{15}$~cm$^{-2}$. 
A scaling of 1.46~kcps/nm$^2$ is naively expected at a dose of $10^{17}$~cm$^{-2}$, which is consistent with Ref.~\onlinecite{Zeng2022}.
This estimation means that a spot size of (10~nm)$^2$ shows a high PL intensity of 146~kcps.
If sufficient PL intensity can be obtained with a hBN flake thinner than the present ones, a three-dimensional spatial resolution of (10~nm)$^3$ is anticipated.
Also, the above result tells us the possibility of even arranging a single V$_\text{B}^-$ defect.

We demonstrated magnetic field imaging with a nanoarray of V$_\text{B}^-$ quantum sensors fabricated by HIM.
Each sensor has a size of (100~nm)$^2$ in hBN with a thickness of less than 100~nm.
The nanometer accuracy of the sensor array position ensured by HIM enables imaging at high spatial resolution.
Moreover, the observed magnetic field sensitivity is sufficient for various physical properties measurements, such as quantum materials.
In particular, as hBN can be incorporated into the interior of vdW devices, we can obtain information about their deep inside, which is unapproachable from the outside.
Optimizing the irradiation conditions of HIM can achieve even higher magnetic field sensitivity and three-dimensional spatial resolution.
The pre-defined position is advantageous for studying systems with slight movements.
Simultaneous measurement of the nanoarray by wide-field microscope will enhance imaging speed~\cite{Sengottuvel2022}.
It is possible to be combined with super-resolution techniques~\cite{Rust2006,Hell1995} to mitigate the restriction of the spacing between sensor spots.
The present achievement opens up applications that take full advantage of the potential of the V$_\text{B}^-$ quantum sensor.

We thank Mr. Tomohiko Iijima (AIST) for the usage of AIST SCR HIM for the helium ion irradiations, Dr. Toshihiko Kanayama (AIST) for helpful discussions since the introduction of HIM at AIST in 2009, Kohei M. Itoh (Keio) for letting us to use the confocal microscope system.
This work was partially supported by ``Advanced Research Infrastructure for Materials and Nanotechnology in Japan (ARIM)" (Proposal Nos. JPMXP1222NM0070 and JPMXP1222UT1131) of the Ministry of Education, Culture, Sports, Science and Technology of Japan (MEXT), ``World Premier International Research Center Initiative on Materials Nanoarchitectonics (WPI-MANA)" supported by MEXT, Daikin Industries, Ltd, and FoPM, WINGS Program, the University of Tokyo.
This work was supported by Grants-in-Aid for Scientific Research (KAKEN) Nos.~JP22K03524, JP19H00656, JP19H05826, JP19H05790, JP23H01103, JP20H00354, and JP22J21412, and Next Generation Artificial Intelligence Research Center at the University of Tokyo.
Correspondence and requests for materials should be addressed to K.S. (kento.sasaki@phys.s.u-tokyo.ac.jp) or K.K. (kensuke@phys.s.u-tokyo.ac.jp).


%

\end{document}